\documentclass[]{spie}  

\usepackage{amsmath,amsfonts,amssymb}
\usepackage{graphicx}
\usepackage[colorlinks=true, allcolors=blue]{hyperref}

\title{Betelgeuse scope: Single-mode-fibers-assisted optical interferometer design for dedicated stellar activity monitoring}

\author[a]{Narsireddy Anugu}
\author[a]{Katie M. Morzinski}
\author[a]{Josh Eisner}
\author[a]{Ewan Douglas}
\author[a]{Dan Marrone}
\author[a]{Steve Ertel}
\author[a]{Sebastiaan Haffert}
\author[a]{Oscar Montoya}
\author[a]{Jordan Stone}
\author[b]{Stefan Kraus}
\author[c]{John Monnier}
\author[d]{Jean-Baptiste Lebouquin}
\author[d]{Jean-Philippe Berger}
\author[e]{Julien Woillez}
\author[f]{Miguel Montarg\`es}

\affil[a]{Steward Observatory, Department of Astronomy, University of Arizona, Tucson, USA}
\affil[b]{School of Physics and Astronomy, University of Exeter,  Exeter, Stocker Road, EX4 4QL, UK}
\affil[c]{University of Michigan, Ann Arbor, MI 48109, USA}
\affil[d]{Institut de Planetologie et d'Astrophysique de Grenoble, Grenoble 38058, France}
\affil[e]{European Southern Observatory, Karl-Schwarzschild-Straße 2, 85748 Garching, Germany}
\affil[f]{Institute of Astronomy, KU Leuven, Celestijnenlaan 200D B2401, 3001 Leuven, Belgium}
\authorinfo{E-mail: nanugu@arizona.edu}

\begin{document} 
\maketitle

\begin{abstract}
Betelgeuse has gone through a sudden shift in its brightness and dimmed mysteriously. This is likely caused by a hot blob of plasma ejected from Betelgeuse and then cooled to obscuring dust. If true, it is a remarkable opportunity to directly witness the formation of dust around a red supergiant star. Today's optical telescope facilities are not optimized for time-evolution monitoring of the Betelgeuse surface, so in this work, we propose a low-cost optical interferometer. The facility will consist of $12 \times 4$~inch optical telescopes mounted on the surface of a large radio dish for interferometric imaging; polarization-maintaining single-mode fibers will carry the coherent beams from the individual optical telescopes to an all-in-one beam combiner. A fast steering mirror assisted fiber injection system guides the flux into fibers. A metrology system senses vibration-induced piston errors in optical fibers, and these errors are corrected using fast-steering delay lines. We will present the design.
\end{abstract}


\section{INTRODUCTION}\label{sec:intro}
\begin{figure}
	\includegraphics[height=0.4\textwidth]{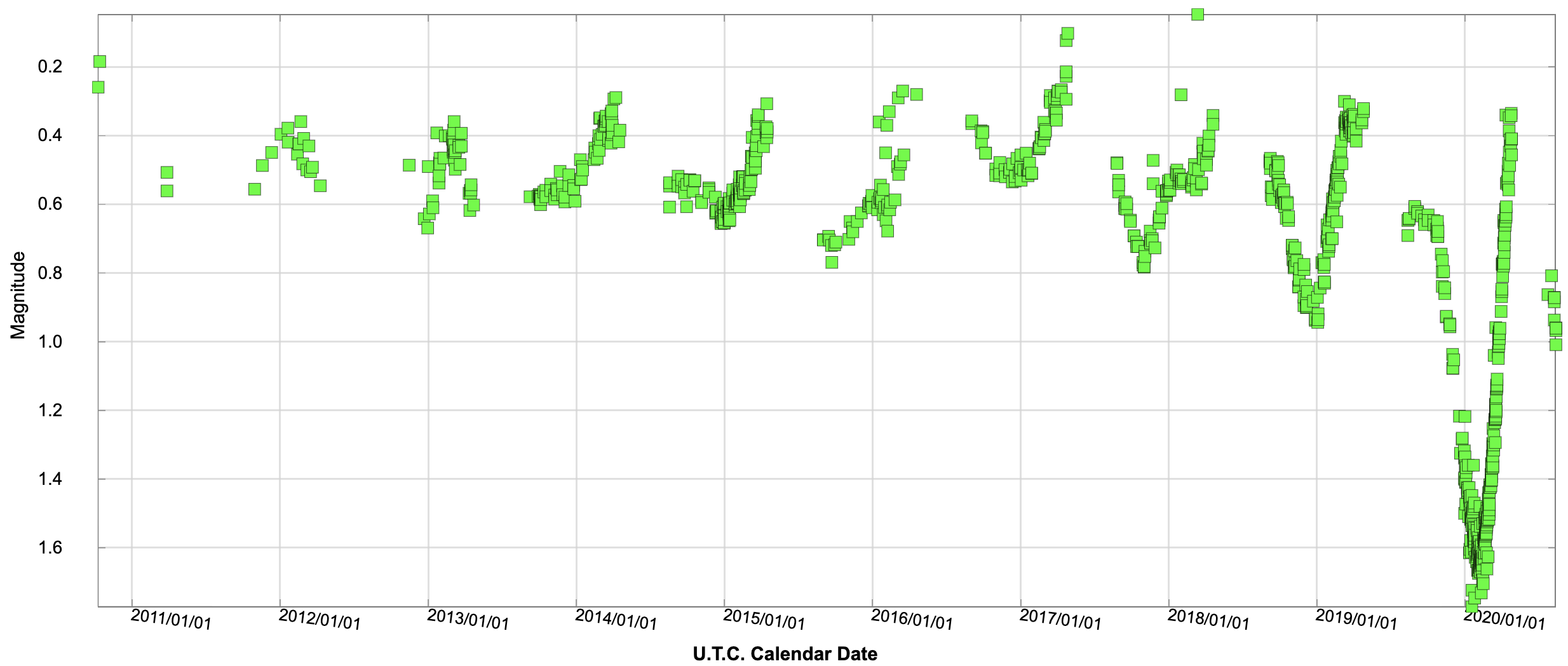}
	\centering
	\caption[a]
	{\label{Fig:1}\small{AAVSO visible light curve of Betelgeuse in V bands in the last ten years. In 2019 November, the star's brightness in the V magnitude started to fall below its average value.  On 2020 February 10, Betelgeuse reached a historical minimum ($1.614 \pm 0.008$) of more than a magnitude dimmer than its average value. At the end of this Great Dimming, Betelgeuse started brightening from 2020 February 10 until the end of 2020 May and since then started dimming again and continuing (last data point 2020 August 15).}}
	
\end{figure}

The bright red supergiant star (RSG), Betelgeuse, is experienced a sudden shift in the brightness in 2019-2020 and has been the subject of much debate. The visual dimming started in 2019 November and reaching a historical minimum ($1.614 \pm 0.008$ magnitude) of only 37\% of its average brightness on 2020 February 10 (see Fig.~\ref{Fig:1}). Subsequently, the optical magnitude began to recover until the end of 2020 May then started dimming again.  These unexpected Great Dimming events pose several questions about its future as it also skipped its normal pulsation cycle of $\sim $ 420 days: (i) Has it entered a new phase of very frequent dimming events? (ii) Being a candidate to become a core-collapse Type II supernova has entered into the pre-supernovae phase?

Observations from various facilities suggest broadly two hypotheses (i) formation of dark spots on its surface, (ii) formation of dust close to the star and blocking light in our line of sight: 

\begin{itemize}
    \item \textbf{Dark spots hypothesis:} Submillimeter observations taken by the James Clerk Maxwell Telescope and Atacama Pathfinder Experiment have claimed that the presence of dark spot(s)\cite{Dharmawardena2020}. 

    \item \textbf{Dust formation and blocking hypothesis:} 
    The spectrophotometric\cite{Levesque2020} and high spectral resolution\cite{Harper2020} observations recorded during the Great Dimming event do not report any notable changes in its effective temperature ($\sim $ 3600 K) or emission lines [Fe II] and [S I] in comparison to past non-dimming observations. 
    
    Differential speckle polarimetry observations show that the polarized flux of the circumstellar envelope around the star is increased after the dimming event (2020 February 10) but remains constant before or during the Great Dimming -- indicating dust formation\cite{Safonov2020}.

    VLT/SPHERE adaptive optics visible polarimetric images\cite{Montarges2020} have shown that the star shape has changed significantly in the Southern Hemisphere and revealed a substantial dimming of the star favoring dust blocking in the line of sight.

    Spatially resolved ultraviolet spectra using the Hubble Space Telescope/Space Telescope Imaging Spectrograph suggest a hot blob of plasma is ejected from the activity of huge convection cells on the Betelgeuse's surface.  This plasma is then cooled to form an enormous cloud of obscuring dust around the Southern hemisphere of star\cite{Dupree2020}. 
    
    All these observations suggest mass ejection from the known large convective cells in the photosphere and then cooled to form the dust cloud in the southern hemisphere in our line of sight.
\end{itemize}

The hypotheses mentioned above insist on high-angular observations of the Betelgeuse photosphere and, importantly, time-evolution monitoring as it entered into a new phase. If the dust-formation-and-blocking theory is correct, this event is a remarkable opportunity to directly witness the formation of dust around a red supergiant star. Where the red supergiant stars are essential contributors to the chemical enrichment of the universe, the mass-loss phase in their evolution is not well understood.  High angular resolution monitoring of the Betelgeuse surface is crucial to spatially probe the photospheric convection and ultimately reveal the mysterious mass-loss processes. Such a program would require a regular -- at least once per week -- a survey of the star's photosphere and circumstellar environment. 

Betelgeuse's angular size is $\sim $  43 mas\cite{Haubois2009,Montarges2016}. Unfortunately, no single, currently operating facility is optimized for such a time-evolution monitoring campaign. The interferometric facilities, COAST\cite{Young2000} and IOTA\cite{Haubois2009}, used as machines of Betelgeuse surface imaging have been retired. Existing monolithic 8-m class apertures, such as LBT, barely resolve the Betelgeuse diameter and thus cannot resolve its convection cells. On the other hand, long-baseline interferometers (ex. existing CHARA\cite{tenBrummelaar2005}, VLTI\cite{Haguenauer2012}, NPOI\cite{Armstrong1998}, and future MROI\cite{Buscher2013}) can resolve the convection cells using its long baselines but suffer from the lack of short baselines, which are essentially needed for detailed image reconstruction\cite{Monnier2003}. The combination of non-redundant masking (NRM) from 8-m class telescopes and long-baseline interferometric observations can prove efficient for the best imaging reconstruction.  NRM allows imaging of the outer-scale overall object shape and while the long-baseline interferometry observations provide complimentary inner-scale convection cell information. However, a tight, extensive, collaborative scheduling between observatories is required for this task is a realistically unfeasible, and regular interval of monitoring observations is very expensive.

Adaptive optics observations of future $>30$~m class facilities such as E-ELT and TMT can image the Betelgeuse convection cells with 5 pixels across the stellar diameter in J-band.  Implementation of non-redundant aperture masking for the first light instruments of E-ELT is being investigated (ex. E-ELT/METIS). However, they will be ready only after 2025, and monitoring observations will be costly.   Doppler imaging technique observations are limited in determining the longitude location of convection cells\cite{Roettenbacher2017}.

Therefore, in this work, we propose a novel, low-cost, $12\times 4$~inch telescopes based optical interferometer design mounted on a maximum baseline $B=12$~m radio dish. This concept can image the surface of Betelgeuse with $\lambda/2B\approx5$~mas angular resolution with $\approx$~8 pixels across the stellar diameter in the $R$-band ($\lambda = 630 - 780$~nm). It will deliver observations of the time-evolution of Betelgeuse's convection cells every night and reveal the phenomena which cause the mass-loss process.

\section{Historical high-angular observations of Betelgeuse}
Betelgeuse's large angular size ($\sim $  43 mas) and brightness ($R$-band apparent magnitude $\sim $  -1) make it a prime target for the red supergiant mass-loss studies. Betelgeuse has been the subject of many studies over the past century, starting from the first angular diameter measurement by Michelson \& Peace in 1920, exactly a century ago\cite{Michelson1921}.  An appropriate historical background of the optical interferometry and recent state-of-the-art of the field can be found elsewhere\cite{Monnier2003}. An asymmetric and multi-feature circumstellar environment of Betelgeuse is detected with infrared wavelength observations using lucky imaging\cite{Kervella2011}.  Recently, its convective activity has been imaged utilizing COAST, IOTA and VLTI optical interferometers\cite{Young2000,Haubois2009,Montarges2016} and spectropolarimetric measurements\cite{Lopez2018}. Advanced numerical hydrodynamic simulations\cite{Chiavassa2010} were conducted and they predict convection cells of various sizes in contrast to observation detected large convection cells\cite{Young2000,Haubois2009,Montarges2016}. Multi-epoch high-angular observations of Betelgeuse photosphere is required to constrain the convection cell size and intensity contrast in the models.

\section{BETELGEUSE SCOPE DESIGN}
\subsection{Concept}
We propose a new facility Betelgeuse scope consists of $12 \times 4$~inch optical Cassegrain telescopes array mounted to the surface of a large radio dish.  Polarization-maintaining single-mode fibers will carry the coherent beams from the individual optical telescopes to the central
beam combining the facility at the control cabin of the radio dish. An all-in-one beam combiner\cite{Monnier2003} (similar to CHARA/MIRC-X\cite{Anugu2020}) will be implemented, and this will allow model-independent aperture synthesis image of Betelgeuse with  $\sim $8 resolution pixels across the Betelgeuse diameter in the $R$-band (630 -- 780 nm) with interferometric angular resolution of $\lambda/2B\approx5$~mas. The radio dish concept is chosen as it relaxes the cost of long delay lines and also the pointing and tracking control for the individual optical telescopes by having a common mount. This is a significant cost-saving concept; most of the interferometer cost goes for the delay lines and the pointing and tracking control of individual telescopes.  Our observing strategy is to point at only one target, Betelgeuse,  the entire possible night, increasing the (u,v) coverage using the Earth's rotation. The benefit of this common mount is that we don't need longer delay lines as in for the traditional non-common mount based long-baseline interferometers, which have to compensate for the changing geometrical delay between wavefronts reaching any two telescopes.

The pointing and tracking errors ($\sim $ 2 arcsec) of mm-wave radio dish are similar to those of optical telescopes\cite{Mangum2006PASP..118.1257M}. We are prepared for technical challenges from atmospheric turbulence, vibrations, and pointing errors from the radio dish in the practical windy conditions. These errors cause difficulties\cite{Woillez2017, Anugu2018} related to the flux injection into single-mode fibers as well as vibration-induced fiber piston errors. Capitalizing on the fact that Beletgeuse is very bright in $R$-band (magnitude $\sim $ -1),  the flux injection into fibers will be maximized using a standard tip-tilt correction with a fast steering mirror and a fast frame rate detector operated in $V$-band. The vibration-induced fiber piston errors will be measured using a state-of-the-art metrology system. It works by sending a coherent laser beam into the single-mode fibers from the beam combiner side and back-reflecting the beam from the telescope side for recording metrology fringes on the same science camera. The metrology laser will measure the instrumental vibration-related piston errors, and the science beam will measure atmospheric piston errors. Piezo-actuated pathlength correctors with a resolution of a micron and a centimeter range will correct both these error sources. The astronomical science measurements will be conducted in the $R$-band, and the metrology system will work at a different wavelength (in $I$-band).

We will use group delay fringe tracking\cite{Monnier2003, Anugu2020} enabled by dispersing the fringes. High-spectral resolution (R $\sim $ 10000) observations will be recorded to allow detection of turbulent motion in the atmosphere by measuring the velocity of the gas at each position\cite{Ohnaka2017}. 

The $R$-band filter has been chosen as it offers the following advantages. 

\begin{figure}
	\includegraphics[height=0.65\textwidth]{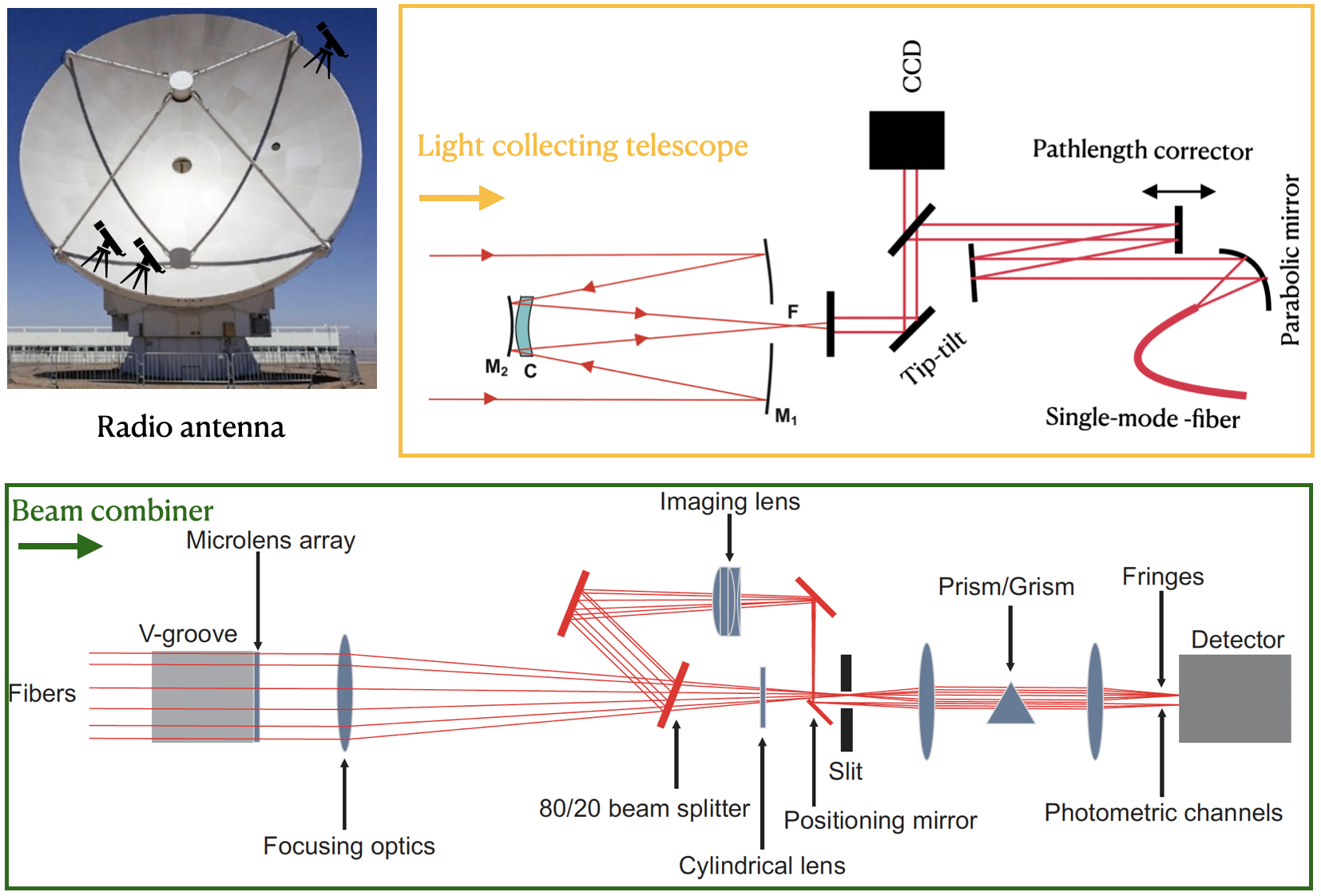}
	\centering
	\caption[a]
	{\label{Fig:2}Cartoon showing the pathfinder Betelgeuse scope concept. Top left: three non-redundant telescope array to be installed on a 6-m radio dish. Yellow box (top right) shows the layout for optical telescope light collection, tip-tilt, and pathlength corrections and fiber light injection at the telescope site. Green box (bottom) shows the layout for the beam combiner instrument, which combines the beams from fibers and makes interferometric fringes. A similar design of CHARA/MIRC-X will be used for the three beam pathfinder Betelgeuse scope\cite{Anugu2020}. Once the telescope beams injected into fibers, they are arranged in a V-groove non-redundantly to allow various spatial frequency fringes. The beams are combined at the focus of the lens in the ``all-in-one" beam combining scheme, and fringes are made. A cylindrical lens compresses the fringe pattern in the spectral direction, and then a prism disperses the light in the perpendicular direction to the spatial direction of fringes. Photometric channels are made with a beam splitter (20 flux split).}

\end{figure}

\begin{figure}
	\includegraphics[height=0.28\textwidth]{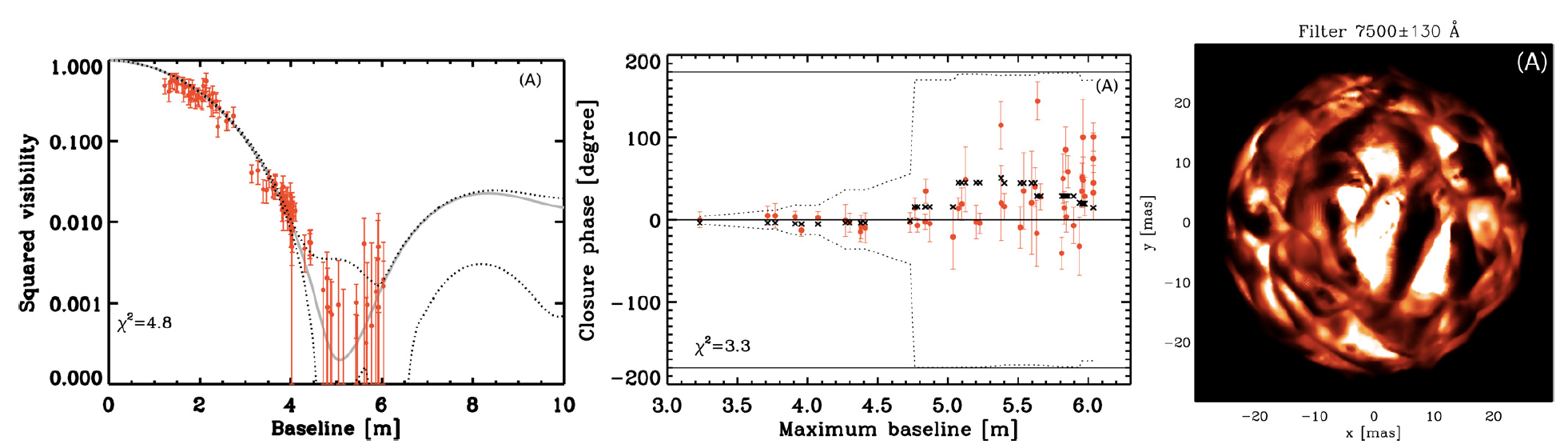}
	\centering
	\caption[a]
	{\label{Fig:3}The visibility curves are plotted in left column, while the closure phases are shown in central column. The right image is the Betelgeuse synthetic image  matched the observation data from Young et al. 2004 from the COAST interferometer. The figure is reproduced from Chiavassa et al. (2010)\cite{Chiavassa2010}.}

\end{figure}

\begin{enumerate}
   \item $ R $-band offers a higher angular resolution than near-infrared wavelengths and is important astrophysically as the dimming is happening in these wavelengths. See Fig.~\ref{Fig:3}, for expected visibilities.
    \item Detectors, which have a fast frame rate, low-readout-noise, and large size, are relatively inexpensive at $ R $-band wavelengths. The fringes of all 12 telescopes (66 baselines) will be recorded on 1024 pixels along the spatial direction and 8 pixels (lowest dispersion mode) along the spectral direction (See Fig.~\ref{Fig:2} for the optical layout). We need the detector with fast frame rates to take short-exposures within the atmospheric coherence time, $\sim500$~Hz. A low-readout-noise enables high-spectral resolution observations, dispersing the light over many pixels. For this $1608 \times 1104$ pixels based C-BLUE ONE camera from First-light-imaging can be a good choice with 660~Hz full window frame rate (for sub-window $1024\times8$~pixels readouts the frame rates even higher), $<3e^-$ readout noise and 70\% quantum efficiency at $R$-band. 
    \item $ R $-band is a trade-off wavelength considering sensitivity  of Betelgeuse (brighter in near-infrared $J=-3$ than $R=-1$) and atmospheric coherence time ($\tau(\lambda) \propto \lambda ^{6/5}$ at most sites it is $\sim10$~ms at $\lambda=500$~nm). Typical atmospheric coherence radius, the Fried parameter $r_0(\lambda) \propto \lambda ^{6/5}$, at $R$-band is approximately equal to $4$-inch, so choosing a $D=4$~inch for the individual telescope size does not require adaptive optics. With a closed-loop tip-tilt correction assisted fiber injection, we estimate to achieve better than 1\% effective throughput, with that we get 250 signal-to-noise ratio at 1~ms coherent integration.
\end{enumerate}
    
We will use single-mode fibers to spatially filter the atmospheric distorted wavefronts to enable stable visibility and closure phase measurements\cite{1998SPIE.3350..856C}. In the visible wavelengths, differential chromatic dispersion is usually large in magnitude if the optical path is not in a vacuum. To compensate this effect, longitudinal dispersion correction systems are built\cite{tenBrummelaar2005}. However, we expect no substantial differential chromatic dispersion for the Betelgeuse scope as a tiny path of light is propagating in the air, so we don't require a longitudinal dispersion correction system. We will use polarization maintaining fibers to maximize the maximum visibility contrast. We will use the same set of light-collecting and fiber injection optics to minimize the polarization differences between individual telescopes to maximize visibility contrast. All the lengths of fibers will be matched closely. If any mismatch in the lengths introduces birefringence, and that can lead to loss of fringe contrast\cite{Anugu2020}.

The Van Cittert-Zernike Theorem relates the source brightness distribution to the contrast of an interferometer's fringes to a unique Fourier frequency\cite{Monnier2003}. The basic interferometric observables are the visibility and closure phase.  Visibility is a measure of the spatial resolution of the source by a given baseline at a given wavelength. The raw visibility is calibrated using calibrator stars with known diameter. The closure phases are related to the asymmetry of source -- the closure phase is null for a centrosymmetric target. The closure phase is, in principle, a self-calibrated measure.  With the $N=12$ telescope interferometer we get $\tfrac{N(N-1)}{2} = 66$ independent visibility and closure phase measurements. 

Ring interferometric configuration is chosen to install the optical telescopes on the edge of the radio dish, which will not interfere with any regular radio observations (Fig.~\ref{Fig:2}). The non-redundant baseline separation decides fringe spatial frequencies of all baselines $u=B/\lambda$~radians$^{-1}$. As aforementioned, the visibility of long-baseline fringes is low in comparison to small baseline fringes. The baselines selection will be made considering Betelgeuse's diameter so that its visibilities do not fall in the zero nulls (Fig.~\ref{Fig:3}). A baseline bootstrapping method based on nearest pairs will be used, in order to track the most resolved baselines efficiently\cite{Monnier2003}.  

Model-independent image reconstruction in the optical interferometry and especially imaging the surface of stars is matured now\cite{Roettenbacher2016,Paladini2018}, with several image reconstruction tools readily available to use, for example, \texttt{macim}\cite{Macim2006}, \texttt{squeeze}\cite{Baron2010} and \texttt{MiRA}\cite{Thiebaut2008}.

\subsection{Advancements to the field} 
Betelgeuse scope offers the following advancements to the field.
\begin{itemize}
    \item The intellectual merit of the project derives from its potential of observing the time-evolution of Betelgeuse's convection cells every night and correlate them to mass-loss phenomena.  In the second half of the year, when Betelgeuse is not observable, the scope will observe the second brightest RSG star, Antares (diameter $\sim $ 37 mas; $R$-band mag $\sim $  -0.65). This target is also interesting astrophysically, where a previous study\cite{Ohnaka2017} suggests its convection alone cannot explain the observed turbulent motions. Betelgeuse scope is also optimized to make observations of very interesting large size symbiotic systems such as R Aqr\cite{Bujarrabal2018A&A...616L...3B}. In R Aqr system, the dwarf, disk, and cloud move in a 44-year orbit around the system’s center of mass. The next eclipse is expected to begin in 2021 and continue until 2024.
    \item The modular aspect of the instrument is that it can be extended to several telescopes array to increase the (u,v)-coverage for better and faster image reconstruction.
    \item The technology demonstration in this proposed project may have significant broader impacts for the future of long-baseline interferometry in using inexpensive optical fibers for light transport. Currently, building kilometer size optical interferometers are limited due to the high cost of vacuum pipes for light transport. Furthermore, small-satellite-enabled interferometric projects may benefit as their size and weight need to be small, where fiber usage helps in place of several mirrors for light transport.  
\end{itemize}

\subsection{Pathfinder on 6-m radio dish}
We are building a pathfinder for a 6-m radio dish owned by the University of Arizona. We have procured one set (out of 12) of light collecting and fiber injection optics and integrating them in our lab at Steward Observatory, Arizona. We need an efficient fiber light injection system as it is the only limitation for throughput.  Note this design has a very few mirror light reflections involved.  

The selection of light collecting and fiber injection hardware is a trade-off in between (i) low-cost-budget requirement and (ii) meeting specifications to maximize flux injection into fibers. 

\paragraph{Telescope:} A smaller tube telescope is preferred for mechanical constraints. A 4~inch telescope from Orion Apex Maksutov-Cassegrain is computed to have a signal-to-noise ratio of 250 at 1 ms integration with an effective throughput of 1\%. 

\paragraph{Tip-tilt actuator:} Speed and large stroke is the requirement to correct large amplitude vibrations or atmospheric tip-tilts.  1-inch mirror from the optics in motion company, OIM101 is selected as it operates at 550~Hz speed (3dB Bandwidth), allows large tip-tilt stroke $\pm1.5^{\circ}$,  with excellent tip-tilt resolution $<2\mu$rad and with good wavefront quality $\lambda/10$ at 633~nm.

\paragraph{Tip-tilt tracking camera:} A large field of CMOS camera is the requirement to detect large amplitude vibrations or atmospheric tip-tilt based offsets.  FLIR BFS-PGE-04S2M-CS CMOS camera is selected, which has $720 \times 540$ pixels with a pixel size of $6.9~\mu$m, operates at full-frame readout at 291 Hz (appropriate ROI will be used) and have quantum efficiency 65\% at $R$-band. It is a GigE camera, and the frame grabbing is done with Ethernet communication. 

\paragraph{Fiber collimator and fiber selection:} Polarization, maintaining single-mode fibers are the requirements to minimize birefringence to allow maximum instrumental visibility contrast. Since we aim to record fringes in narrow-bandwidth filter off-axis parabola based fiber coupling is not the main requirement. We selected a readily available Thorlabs air-spaced doublet fiber collimator. Polarization maintained single-mode fibers (Nufern PM-S630-HP, single-mode 630 -- 780 nm).

\paragraph{Real-time control software:} In total, two active control loops run during the observations: 
\begin{enumerate}
    \item Tip-tilt tracking to maximize flux injection into single-mode fibers.
    \item Fringe tracking to measure OPD offsets caused by the atmosphere at science wavelengths and  OPD offsets caused by the telescope and instrument vibrations at metrology wavelengths.
\end{enumerate}

We are developing the tip-tilt tracking software consists of following software modules:
\begin{enumerate}
    \item Reading frames from the tip-tilt tracking camera
    \item Computing the tip-tilt offsets
    \item Guiding the tip-tilt actuator for an efficient flux injection into the fiber
\end{enumerate}

For the real-time software, we have adapted CHARA Array software architecture\cite{tenBrummelaar2005}: (i) C-written servers to communicate with hardware and (ii) C-written GNOME Tool Kit (GTK) based Graphical User Interface (GUI) clients to communicate with the servers during the observations. We will adapt CHARA/MIRC-X real-time control software for data processing and group delay tracking and data recording sequences\cite{Anugu2018a,Anugu2020}. We will adapt already tested and used 6-telescope CHARA/MIRC-X  data reduction pipeline written in python\cite{Anugu2020}.

We are delayed by the  COVID-19 pandemic and anticipate to install this light collecting optics in 2020 September and test the throughput into the fibers. We expect installing the three-beam combiner once the tests finished. For the beam combiner, we will use CCD-39 camera ($80\times80$ pixels with each 24 $\mu$m,  readout noise $7e^-$/px, frame rate $> 1$~KHz) from the Steward LBTI lab. 

Fig.~\ref{Fig:3} shows the expected visibilities, closure phases for $B=6$~m and also a data matched synthetic image match similar baselines of Betelgeuse scope in the $R$-band.

\section{CONCLUSION}
Betelgeuse is a red supergiant star and candidate to become a core-collapse Type II supernova. The Great Dimming of Betelgeuse attracted enormous public attention. Betelgeuse convection cell activity thought is a reason for this dimming by forming a clump of dust in our line of sight. High angular resolution measurements are crucial to spatially probe the photospheric convection and ultimately reveal the mysterious mass-loss processes. 

A dedicated interferometric facility such as the proposed Betelgeuse scope with $12 \times 4$ telescopes resulted in 66 visibilities, and 66 independent closure phases allows taking snapshot images of convection cells and time-evolution monitoring. This is a design leveraging advances in the recent optical interferometry field and in the telecommunication industry for astronomy. Our efforts for building a prototype on a 6-m radio dish are underway, and we anticipate installing it at the end of 2020.

\section{Acknowledgments}
Anugu acknowledges support from the Steward Observatory Fellowship in Instrumentation and Technology Development. We thank you the variable star observations from the AAVSO International Database contributed by observers worldwide and used in this research.

\bibliography{report} 
\bibliographystyle{spiebib} 

\end{document}